\documentclass[preprint,aps,12pt,showpacs,nofootinbib,tightenlines]{revtex4}
\usepackage{mathrsfs}
\usepackage{amsmath}
\usepackage{amssymb}
\usepackage{epsfig}
\usepackage{graphicx}
\def\be{\begin{eqnarray}}
\def\en{\end{eqnarray}}
\def\non{\nonumber\\}

\def\prd{{Phys. Rev. D}~}
\def\prl{{ Phys. Rev. Lett.}~}
\def\plb{{ Phys. Lett. B}~}

\def\epjc{{ Eur. Phys. J. C}~}

\begin{document}
\title{The decays $B\to \Psi(2S)\pi(K),\eta_c(2S)\pi(K)$ in the pQCD approach beyond the leading-order}
\author{Zhi-Qing Zhang
\footnote{Electronic address: zhangzhiqing@haut.edu.cn}} 
\affiliation{\it \small  Department of Physics, Henan University of
Technology, Zhengzhou, Henan 450052, China } 
\date{\today}
\begin{abstract}
Two body $B$ meson decays involving the radially excited meson $\psi(2S)/\eta_c(2S)$ in the final states are
studied by using the perturbative QCD (pQCD) approach. We find that: (a) The branching ratios for the decays involving $K$
meson are predicted as $Br(B^+\to\psi(2S)K^+)=(5.37^{+1.85}_{-2.22})\times10^{-4},
Br(B^0\to\psi(2S)K^0)=(4.98^{+1.71}_{-2.06})\times10^{-4}, Br(B^+\to\eta_c(2S)K^+)=(3.54^{+3.18}_{-3.09})\times10^{-4}$, which are
consistent well with the present data when including the next-to-leading-order (NLO) effects. Here the NLO effects are
from the vertex corrections and the NLO Wilson coefficients. The large errors in the decay $B^+\to\eta_c(2S)K^+$ are mainly
induced by using the decay constant $f_{\eta_c(2S)}=0.243^{+0.079}_{-0.111}$ GeV with large uncertainties. (b) While there seems
to be some room left for other higher order corrections or the non-perturbative long distance contributions in the decays
involving $\pi$ meson, $Br(B^+\to\psi(2S)\pi^+)=(1.17^{+0.42}_{-0.50})\times10^{-5},
Br(B^0\to\psi(2S)\pi^0)=0.54^{+0.20}_{-0.23}\times10^{-5}$, which are smaller than the present data. The results for other decays can be tested at the running
LHCb and forthcoming Super-B experiments. (c) There is no obvious evidence of the direct CP violation being seen in the decays
$B\to \psi(2S)\pi(K), \eta_c(2S)\pi(K)$ in the present experiments,
which is supported by our calculations. If a few percent value is confirmed in the future
, it would indicate new physics definitely.
\end{abstract}

\vspace{1cm}

\maketitle


\section{Introduction}\label{intro}
It is well known that $\eta_c(2S)$ and $\Psi(2S)$ are the first radially
excited states of the S-wave ground states $\eta_c(1S)$ and $J/\Psi$,
respectively. These two excited charmonia states have been observed
in the $B$ meson decays in the experiments \cite{pdg14},
\be
Br(B^+\to \eta_c(2S)K^+)=(3.4\pm1.8)\times10^{-4},\\
Br(B^+\to \psi(2S)K^+)=(6.27\pm0.24)\times10^{-4},\\
Br(B^+\to \psi(2S)\pi^+)=(2.44\pm0.30)\times10^{-5},\\
Br(B^0\to \psi(2S)K^0)=(6.2\pm0.5)\times10^{-4}. \en
Furthermore, the direct CP-violating  asymmetries of the two charged decay channels
are given by PDG \cite{pdg14}, though with large uncertainties:
\be
A_{CP}(B^{+}\to \psi(2S)K^+)&=&(-2.4\pm2.3)\%,\\
A_{CP}(B^{+}\to \psi(2S)\pi^+)&=&(3.0\pm6.0)\%. \en

In fact, $B$ meson exclusive decays into charmonia have been received a lot of
attentions for many years. They are regarded as the golden channels
in researching CP violation and exploring new physics.
At the same time, they play the important roles in
testing the unitarity of the Cabibbo-Kobayashi-Maskawa (CKM)
triangle.  Moreover, these decays
are ideal modes to check the different factorization approaches, such
as the naive factorization assumption (FA) \cite{ali}, QCD-improved
factorization (QCDF) \cite{beneke,beneke2}, light-cone sum rules (LCSR)
\cite{melic}, and the perturbtive QCD (pQCD) approach \cite{chen}.
Most of these approaches can not work well when describing these decays, such as
$B\to (J/\Psi,\eta_c,\chi_{c0},\chi_{c1})K^{(*)}$.
Because their predictions about the branching ratios are too small to explain
the data. For example, the branching ratios of the decays $B^{+}\to
J/\Psi K^+$ and $B^{0}\to J/\Psi K^0$ measured by Babar were about
$(10.61\pm0.50)\times10^{-4}$ and $(8.69\pm0.37)\times10^{-4}$
\cite{babar}, respectively, which imply a larger Wilson coefficient
$a_2\approx0.20\sim0.30$ \cite{cheng}. So the naive factorization
assumption does not work with the Wilson coefficient $a_2\sim 0$.
As for the QCDF approach, it has
been found that the end point singularities exist in the amplitudes
from the hard spectator scattering and annihilation diagrams, where
the contributions
are parametrized into \be
\ln\frac{m_B}{\Lambda}(1+\rho_He^{i\delta_H}),
\;\;\;\;\;\ln\frac{m_B}{\Lambda}(1+\rho_Ae^{i\delta_A}),
\label{para} \en respectively. These non-universal and
uncontrollable parameters $\rho_{H,A}$ and $\delta_{H,A}$ will
induce large theoretical uncertainties. What's worse, when the
emitted meson is heavy, such as $D$ and charmonia states $J/\Psi, \Psi(2S),
\eta_c(2S)$, the QCDF will break down. For example, in the $\bar
B^0\to D^0\pi^0$ decay, since the $D^0$ meson is not a compact
object with small transverse extension, it will strongly interact
with the $B\pi$ system, which makes the factorization fail.
Fortunately, the transverse size of the charmonium is small in the
heavy quark limit, so the authors of Ref.\cite{chay} considered that
the QCDF method could be used to the decay $B\to J/\Psi K$, while they
found that the leading-twist (twist-2) contributions were too small
to explain the data. Then the authors of Ref.\cite{cheng} calculated the
twist-3 contributions, where the divergent integral was
parametrized as Eq.(\ref{para}). But they still could not explain
the experimental data. The end-point singularities were also found
in the twist-3 amplitudes for the decays $B\to \eta_c(1S) K,
\eta_c(2S) K$ \cite{song}, where the QCDF approach was used. The
LCSR approach was also insufficient to account for the data of these
B meson decays into charmonia \cite{melic}.

While under the pQCD approach,
the spectator quark and other quarks are connected by
one hard gluon (shown in Fig.1). Unlike the QCDF approach, the hard part of the pQCD approach consists of six quarks
rather than four. Certainly, there also exist the soft and collinear gluon exchanges between quarks.
So the double logarithms $\ln^2{Pb}$ will arise from the overlap of the soft and collinear divergences
in radiative corrections to the meson wave functions, $P$ being the dominant light-cone component of a meson momentum, $b$
being the conjugate variable of parton transverse momentum $k_{T}$. One can use the $k_T$ resummation \cite{lihn} to organize
these leading double logarithms for all loop diagrams into a Sudakov factor, which
suppresses the long distance contributions in the large $b$ region. When the end-point region with a momentum fraction
$x\to 0,1$ is important for the hard amplitude,
the corresponding large double logarithms $\alpha_s \ln^2x$ shall appear in the hard amplitude. One can
use the threshold resummation \cite{lihn2} to organize this type of double logarithms for all loop diagrams into a jet function,
which suppresses the end-point behavior of the hard amplitude.
With the Sudakov factor and the jet function, one can evaluate all possible Feynman diagrams
for the six-quark amplitude directly, including the nonfactorizable emission diagrams and annihilation type diagrams.
But it is difficult to calculate these two kinds of contributions in QCDF approach. The pQCD approach has been used
to B meson decays into charmonia
such as $B\to J/\Psi K^{(*)}$ in Refs.\cite{chen,li0}, where the consistent results with
the experimental data were obtained.

So we would like
to use the pQCD approach to study $B\to \psi(2S)P,\eta_c(2S)P$ (here $P$ refers to the
pseudo-scalar meson $\pi$ or $K$) decays. Except the full
leading-order (LO) contributions, the next-to-leading-order (NLO)
contributions, which are mainly from the NLO Wilson coefficients and
the vertex corrections to the hard kernel, are also included.
Certainly, other NLO contributions, such as the quark loops and the
magnetic penguin corrections, are available in the
literatures \cite{li,xiao}, but they will not contribute to these
considered decays.

We review the LO predictions for the $B\to \psi(2S)P,\eta_c(2S)P$ decays including those for the main NLO contributions, in Sec. II. We perform
the numerical study in Section IV, where the theoretical uncertainties are also considered. Section V is the
conclusion. It is noticed that we will use the abbreviation $\psi$ and $\eta_c$ to denote the mesons $\psi(2S)$ and
$\eta_c(2S)$ unless specified in the following.
\section{the Leading-Order Predictions and the main next-to-leading order corrections}
As previously stated, the pQCD factorization approach has been used to calculate many two-body charmed B meson
decays, and has obtained consistent results with experiments. So we use this approach to consider the decays
$B\to \psi(2S)\pi(K),\eta_c(2S)\pi(K)$, and the corresponding effective Hamiltonian can be written as:
\be
\emph{H}_{eff}=\frac{G_F}{\sqrt2}\left[V^*_{cb}V_{cq'}(C_1(\mu)O^c_1(\mu)+C_2(\mu)O^c_2(\mu))-
V^{*}_{tb}V_{tq'}\sum^{10}_{i=3}C_i(\mu)O_i(\mu)\right],
\en
where $C_i(\mu)$ are Wilson coefficients at the renormalization scale $\mu$,
$q'=d$ ($q'=s$) for $b\to d$ ($b\to s$) transition. $V$ represents for the Cabibbo-Kobayashi-Maskawa (CKM) matrix
element, and the four fermion
operators $O_i$ are given as:
\be
O^c_1&=&(\bar
q'_{i}c_{j})_{V-A}(\bar c_{j}b_{i})_{V-A},\;\;\;
O^c_2=(\bar q'_{i}c_{i})_{V-A}(\bar
c_{j}b_{j})_{V-A},\\
O_3&=&(\bar
q'_{i}b_{i})_{V-A}(\bar q_{j}q_{j})_{V-A},\;\;\;
O_4=(\bar q'_{i}b_{j})_{V-A}(\bar
q_{j}q_{i})_{V-A},\\
O_5&=&(\bar
q'_{i}b_{i})_{V-A}(\bar q_{j}q_{j})_{V+A},\;\;\;
O_6=(\bar q'_{i}b_{j})_{V-A}(\bar
q_{j}q_{i})_{V+A},\\
O_7&=&\frac{3}{2}(\bar
q'_{i}b_{i})_{V-A}\sum_{q}e_{q}(\bar q_{j}q_{j})_{V+A},\;\;\;
O_8=\frac{3}{2}(\bar q'_{i}b_{j})_{V-A}\sum_{q}e_{q}(\bar
q_{j}q_{i})_{V+A},\\
O_9&=&\frac{3}{2}(\bar
q'_{i}b_{i})_{V-A}\sum_{q}e_{q}(\bar q_{j}q_{j})_{V-A},\;\;\;
O_{10}=\frac{3}{2}(\bar q'_{i}b_{j})_{V-A}\sum_{q}e_{q}(\bar
q_{j}q_{i})_{V-A},
\en
with $i,j$ being the color indices.
\begin{figure}[t]
\vspace{-4cm} \centerline{\epsfxsize=18 cm \epsffile{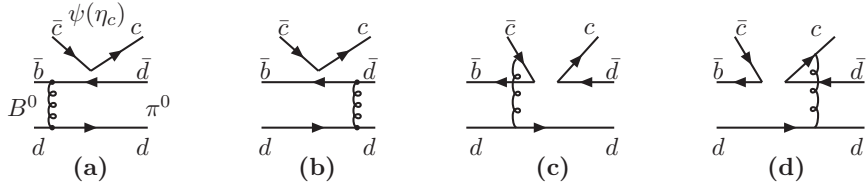}}
\vspace{-18.8cm} \caption{ Feynman Diagrams contributing to the decay $B^0\to \psi(\eta_c)\pi^0$ at leading order.
By replacing the spectator quark $d$ with $s$ quark, one can obtain the Feynman Diagrams for the decay $B^0\to \psi(\eta_c)K^0$. }
 \label{Figure1}
\end{figure}
At the leading order, the relevant Feynman diagrams only include the factorizable and
non-factorizable emission diagrams, as shown in Fig.1, where we take the decays $B^0\to \psi(\eta_{c})\pi^0$ as examples. If the emission particle is the
vector meson $\psi(2S)$, then the amplitude for the factorizable emission diagrams Fig.1(a) and Fig.1(b)
can be written as:
\be
F^{V-A}_{\psi P}&=&8\pi C_F m^4_Bf_{\psi}\int_0^1 dx_1 dx_3
\int_0^\infty b_1 db_1 b_3 db_3 \phi_B(x_1,b_1)\left\{\left[
(1-r^2_\psi)(1+(1-r^2_{\psi})x_3)\right.\right.\non &&\left.\left.\left.\times\phi^A_P(x_3)-2x_3r_P(1-r_\psi^2)
(\phi^P_P(x_3)+\phi^T_P(x_3))+r_P((1-r_\psi^2)\phi^P_P(x_3)\right.\right.\right.
\non &&\left.\left.\left.
+(1+r^2_\psi)\phi^T_P(x_3)\right)\right]\alpha_s(t_a)E_e(t_a)h_e(x_1,x_3,b_1,b_3)S_t(x_3)\right.\non &&\left.
+\alpha_s(t_b)E_e(t_b)h_e(x_3,x_1,b_3,b_1)S_t(x_1)2r_P(1-r_\psi^2-x_1)\phi^P_P(x_3)
\right\},
\label{ab0}
\en
where $r_\psi=m_\psi/m_B,r_P=m^P_0/m_B(P=\pi,K$), the color factor $C_F=4/3$, and
$\phi^{A(P,T)}_{P(P=\pi,K)}$ are the twist-2 (twist-3) distribution amplitudes for the meson $\pi$ or $K$.
$x_1$ and $x_3$ are the light quark momentum fractions in the $B$ and $P$ mesons, respectively. The evolution factors evolving
the Sudakov factors and the jet function $S_t(x)$ are listed as:
\be
E_e(t)=\alpha_s(t)\exp[-S_B(t)-S_{\pi}(t)], \;\;\;\;
S_t(x)=\frac{2^{1+2c}\Gamma(3/2+c)}{\sqrt{\pi}\Gamma(1+c)}[x(1-x)]^c,
\label{suda1}
\en
with $c=0.3$, and $S_B(t), S_{\pi}(t)$ being the Sudakov factors, which can be found in Ref.\cite{zhang}.
The hard functions $h_{e}$ is given as:
\be
h_e(x_1,x_3,b_1,b_3)&=&K_0(\sqrt{x_1x_3(1-r^2_\psi)}m_Bb_1)\left[\theta(b_1-b_3)K_0(\sqrt{x_3(1-r^2_\psi)}m_Bb_1)\right.\non &&\left.
I_0(\sqrt{x_3(1-r^2_\psi)}m_Bb_3)+\theta(b_3-b_1)K_0(\sqrt{x_3(1-r^2_\psi)}m_Bb_3)
\right.
\non && \left.I_0(\sqrt{x_3(1-r^2_\psi)}m_Bb_1)\right].
\en

The amplitude for the non-factorizable spectator diagrams Fig.1(c) and Fig.1(d) is given as:
\be
M^{V-A}_{\psi P}&=&\frac{32}{\sqrt6}\pi C_F m^4_B \int_0^1 dx_1 dx_2 dx_3
\int_0^\infty b_1 db_1 b_2 db_2 \phi_B(x_1,b_1)\non
&&\times\left\{\left[(r^2_\psi(2-r_\psi^2)(x_3-x_2)-x_3)\phi_P^A(x_3)+2r_3(r^2_\psi(2x_2-x_3)+x_3)\phi^T_P(x_3)\right]
\right.\non &&\left.\times \psi^L(x_2,b_2)+2r_cr_\psi\left[(1-r^2_\psi)\phi^A_P(x_3)-2r_P\phi^T_P(x_3)\right]\psi^t(x_2,b_2)\right\}
\non && \times \alpha_s(t)E_d(t)h_d(x_1,x_2,x_3,b_1,b_2),
\label{cd0}
\en
where the twist-2(twist-3) distribution amplitudes $\psi^{L(t)}(x_2,b_2)$ can be found in the next section. $x_2$ is
the $c$ quark momentum fraction in $\psi(2S)$ meson.
The evolution factor $E_d(t)$ and
the hard function $h_d$ are
listed as: \be
E_{d}(t)&=&\alpha_s(t)\exp[-S_B(t)-S_{\pi}(t)-S_\psi(t)|_{b_1=b_2}],\label{edd}\\
h_{d}(x_1,x_2,x_3,b_1,b_2)&=&\left[\theta(b_1-b_2)K_0(\sqrt{x_1x_3(1-r^2_\psi)}m_Bb_1)
I_0(\sqrt{x_1x_3(1-r^2_\psi)}m_Bb_2)\right.\non && \left.+(b_1\leftrightarrow b_2)\right]
\left(\begin{matrix}K_0(A_dm_Bb_2)& \text{for} A^2_d\geq 0\\
\frac{i\pi}{2}H^{(1)}_0(\sqrt{|A^2_d|}m_Bb_2)&\text{for} A^2_d\leq
0\\\end{matrix}\right), \label{hdd}\en with the variable $A^2_d=r_c^2+(x_1-x_2)(x_2r^2_\psi+x_3(1-r^2_\psi))$, and $K_0, I_0$ and
$H_0$ being the modified Bessel functions.

If the emission particle is the
pseudo-scalar meson $\eta_c(2S)$, then the corresponding amplitude $F^{V-A}_{\eta_c P}$ can be obtained from
$F^{V-A}_{\psi P}$ by the replacements of the parameters $f_{\psi}$ and $r_\psi$ with $f_{\eta_c}$ and $r_{\eta_c}$,
respectively. While there are many differences between the nonfactorizable spectator amplitudes
$M^{V-A}_{\eta_c P}$ and
$M^{V-A}_{\psi P}$ because of the different Lorentz structures between the wave functions of $\eta_c(2S)$ and $\psi(2S)$. Here
$M^{V-A}_{\eta_c P}$ is listed as following:
\be
M^{V-A}_{\eta_c P}&=&\frac{32}{\sqrt6}\pi C_F m^4_B \int_0^1 dx_1 dx_2 dx_3
\int_0^\infty b_1 db_1 b_2 db_2 \phi_B(x_1,b_1)\non
&&\times(1-r^2_{\eta_c})x_3\left[(r^2_{\eta_c}-1)\phi^A_P(x_3)+2r_P\phi_P^T(x_3)\right]\psi^v(x_2,b_2)
\non && \times \alpha_s(t)E_d(t)h_d(x_1,x_2,x_3,b_1,b_2),
\label{cd1}
\en
where $r_{\eta_c}=m_{\eta_c}/m_B$, and the twist-3 distribution amplitudes of $\eta_c$ meson do not
contribute to
the amplitude. It is different from the case of $M^{V-A}_{\psi P}$. While the evolution factor $E_d(t)$ and
the hard function $h_d$ are similar with those given in Eq.(\ref{edd}) and Eq.(\ref{hdd}).

By combining the amplitudes from the different Feynman diagrams, one can get the total decay amplitude for the decays
$B\to \psi(\eta_c)\pi$:
\be
\emph{M}(B\to \psi(\eta_c)\pi)&=&F^{V-A}_{\psi(\eta_c) \pi}\left[V^*_{cb}V_{cd}a_2-V^*_{tb}V_{td}(a_3+a_5+a_7+a_9)\right]\non &&
+M^{V-A}_{\psi(\eta_c) \pi}\left[V^*_{cb}V_{cd}C_2-V^*_{tb}V_{td}(C_4-C_6-C_8+C_{10})\right],
\label{totalam}
\en
where $V_{ij}$ is the CKM matrix element and the combinations of Wilson coefficients $a_2=C_1+C_2/3, a_i=C_i+C_{i+1}/3$ with $i=3,5,7,9$.
The amplitudes $F^{V-A}_{\psi(\eta_c)\pi}$ and $M^{V-A}_{\psi(\eta_c) \pi}$
are given in Eqs.(\ref{ab0}) and (\ref{cd0}), respectively.  As for the decays
$B\to \psi(\eta_c)K$, the total amplitude can be obtained by replacing $F^{V-A}_{\psi(\eta_c) \pi}, M^{V-A}_{\psi(\eta_c) \pi}, V_{cd}$ and $V_{td}$
with $F^{V-A}_{\psi(\eta_c) K}, M^{V-A}_{\psi(\eta_c) K}, V_{cs}$ and $V_{ts}$, respectively in Eq.(\ref{totalam}).

As stated before, the NLO corrections to the hard kernel for the
decays $B\to \psi(2S) P$ and $B\to \eta_c(2S) P$ are simpler compared with
other B meson decays such as $B\to \pi K, \rho K$. Here only the
vertex corrections are need to be considered. Since the vertex
corrections can reduce the dependence of the Wilson coefficients on
the renormalization scale $\mu$, they play the important roles in
the NLO analysis. It is well known that the nonfactorizable
contributions are small \cite{li1}, we concentrate only on the
vertex corrections to the factorizable amplitudes, as shown in
Fig.2. Furthermore, the infrared divergences from the soft
and collinear gluons in these Feynman diagrams can be canceled each
other. That is to say, these corrections are free from the end-point
singularity in collinear factorization theorem, so we can simply
quote the QCDF expressions for the vertex corrections: their effects
can be combined into the Wilson coefficients, \be
a_2&\rightarrow& a_2+\frac{\alpha_s C_F}{4\pi N_c}C_2(-18+12\ln\frac{m_b}{\mu}+f_I),\\
a_i&\rightarrow& a_{i}+\frac{\alpha_s C_F}{4\pi N_c}C_{i+1}(-18+12\ln\frac{m_b}{\mu}+f_I), (i=3,9),\\
a_i&\rightarrow& a_{i}+\frac{\alpha_s C_F}{4\pi N_c}C_{i+1}(6-12\ln\frac{m_b}{\mu}-f_I), (i=5,6),
\en
with the function $f_I$ defined as \cite{liu}:
\be
f_I&=&\frac{2\sqrt{2N_C}}{f_{\psi(\eta_c)}}\int^1_0 dx_2\psi^{L(v)}(x_2)\left[\frac{3(1-2x_2)}{1-x_2}\ln x_2-3\pi i\right.\non &&\left.
+3\ln(1-r^2_{\psi(\eta_c)})+\frac{2r^2_{\psi(\eta_c)}(1-x_2)}{1-r^2_{\psi(\eta_c)}x_2}\right],
\en
where we have neglected the terms proportional to $r^4_{\psi(\eta_c)}$. Certainly, in the following numerical
analysis, the NLO Wilson coefficients will be used in the NLO calculations.
\begin{figure}[]
\vspace{-4cm} \centerline{\epsfxsize=18 cm \epsffile{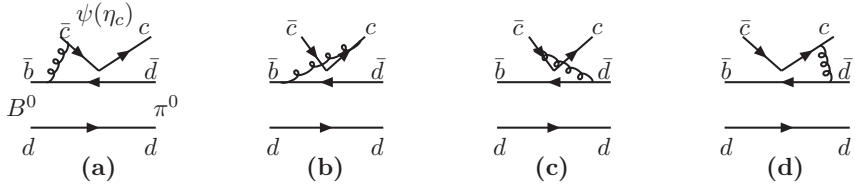}}
\vspace{-18.8cm} \caption{ NLO vertex corrections to the
factorizable amplitudes for the decays $B^0\to \psi(\eta_c)\pi^0$.}
 \label{Figure1}
\end{figure}
\section{Numerical results and discussions} \label{numer}
We use the following input parameters for the numerical calculations
\cite{pdg14,zhou}: \be
f_B&=&210 \text{MeV},  f_{\psi(2S)}=0.296^{+0.003}_{-0.002} \text{GeV}, f_{\eta_c(2S)}=0.243^{+0.079}_{-0.111} \text{GeV},\\
M_B&=&5.28 \text{GeV}, M_{\psi(2S)}=3.686 \text{GeV}, M_{\eta_c(2S)}=3.639 \text{GeV},\\
M_W&=&80.41 \text{GeV}, \tau_B^\pm=1.638\times 10^{-12}
\text{s},\tau_{B^0}=1.519\times 10^{-12} \text{s}. \en
For the CKM
matrix elements, we adopt the Wolfenstein parametrization and the
updated values $A=0.814, \lambda=0.22537, \bar\rho=0.117\pm0.021$
and $\bar\eta=0.353\pm0.013$ \cite{pdg14}. With the total
amplitudes, one can write the decay width as: \be
\Gamma(B\to\psi(\eta_c)P)=\frac{G^2_F}{32\pi
m_B}(1-r^2_{\psi(\eta_c)})|M(B\to\psi(\eta_c)P)|^2. \en The wave
functions of $B, \pi$ and $K$ have been well defined in many works,
while those of the two excited charmonia states exist many
uncertainties. Here we take the harmonic-oscillator model
\cite{zhou}: \be
\psi^{L,v}(x,b)&=&\frac{f_{2S}}{2\sqrt{2N_c}}N^{L,v}x(1-x)\mathcal{T}(x)e^{-x(1-x)\frac{m_c}{\omega}[\omega^2b^2+(\frac{2x-1}{2x(1-x)})^2]},\\
\psi^{t}(x,b)&=&\frac{f_{2S}}{2\sqrt{2N_c}}N^{t}(2x-1)^2\mathcal{T}(x)e^{-x(1-x)\frac{m_c}{\omega}[\omega^2b^2+(\frac{2x-1}{2x(1-x)})^2]},
\en
with
\be
\mathcal{T}(x)=1-4b^2m_c\omega x(1-x)+\frac{m_c(2x-1)^2}{\omega x(1-x)},
\en
where $f_{2S}$ refers to the decay constant $f_{\psi}$ or $f_{\eta_c}$, the free parameter $\omega=0.2\pm0.1$ GeV and the $c$ quark mass
$m_c=1.275\pm0.025$ GeV. The main errors come from these parameters and $\omega_b=0.4\pm0.1$ for $B$ meson wave funcion
in our calculations.

\begin{table}
\caption{Our LO and NLO predictions of the branching ratios for the decays $B\to\psi(\eta_c)K, \psi(\eta_c)\pi$. The fist
uncertainty comes from the $\omega_b=0.4\pm0.1$ for $B$ meson, the second and the third uncertainties are from the free
parameter $\omega=0.2\pm0.1$ and the $c$ quark mass $1.275\pm0.025$ GeV for $\psi(2S)/\eta_c(2S)$ meson. The last one comes from
the decay constant $f_{\psi(2S)}=0.296^{+0.003}_{-0.002}$ GeV/$f_{\eta_c(2S)}=0.243^{+0.079}_{-0.111}$ GeV.
The data are listed in the second column \cite{pdg14}.}
\begin{center}
\begin{tabular}{c|c|c|c}
\hline\hline mode& data &LO&NLO \\
\hline
$B^+\to\psi(2S)K^+(\times10^{-4})$&$6.27\pm0.24$&$2.39^{+1.11+0.18+0.04+0.05}_{-0.81-0.31-0.03-0.03}$&$5.37^{+1.61+0.90+0.04+0.11}_{-1.17-1.89-0.03-0.07}$ \\
\hline
$B^0\to\psi(2S)K^0(\times10^{-4})$&$6.2\pm0.5$&$2.22^{+1.00+0.16+0.03+0.04}_{-0.67-0.29-0.03-0.02}$&$4.98^{+1.49+0.84+0.04+0.10}_{-1.09-1.75-0.03-0.06}$\\
\hline
$B^+\to\psi(2S)\pi^+(\times10^{-5})$&$2.44\pm0.30$&$0.47^{+0.24+0.04+0.01+0.01}_{-0.15-0.04-0.00-0.00}$&$1.17^{+0.36+0.22+0.02+0.03}_{-0.25-0.43-0.01-0.01}$\\
\hline
$B^0\to\psi(2S)\pi^0(\times10^{-5})$&$1.17\pm0.17\pm0.08$&$0.22^{+0.11+0.03+0.01+0.02}_{-0.07-0.02-0.00-0.00}$&$0.54^{+0.17+0.10+0.01+0.02}_{-0.12-0.20-0.01-0.00}$\\
\hline
$B^+\to\eta_c(2S)K^+(\times10^{-4})$&$3.4\pm1.8$&$2.33^{+1.02+0.23+0.02+1.76}_{-0.69-0.17-0.02-1.64}$&$3.54^{+1.23+1.18+0.05+2.68}_{-0.87-1.62-0.04-2.49}$\\
\hline
$B^0\to\eta_c(2S)K^0(\times10^{-4})$&$--$&$2.16^{+0.95+0.21+0.02+1.63}_{-0.64-0.16-0.02-1.52}$&$3.29^{+1.13+1.08+0.04+2.48}_{-0.81-1.51-0.05-2.32}$\\
\hline
$B^+\to\eta_c(2S)\pi^+(\times10^{-6})$&$--$&$5.82^{+2.51+0.70+0.06+4.40}_{-1.69-0.49-0.06-4.10}$&$9.03^{+3.05+3.08+0.11+6.82}_{-2.20-2.56-0.11-6.37}$\\
\hline
$B^0\to\eta_c(2S)\pi^0(\times10^{-6})$&$--$&$2.72^{+1.18+0.33+0.03+2.06}_{-0.79-0.23-0.02-1.92}$&$4.19^{+1.41+1.43+0.06+3.16}_{-1.02-1.97-0.05-2.96}$\\
\hline\hline
\end{tabular}\label{tab1}
\end{center}
\end{table}

Using the input parameters and the wave functions as specified in
this section, we give the LO and NLO predictions for the considered
decays in Table I. For these color-suppressed decays, the amplitudes
associated with the Wilson coefficient $C_1+C_2/3$ usually play the
dominant roles. It is instructive to check the contributions from
these amplitudes at the leading-order and the next-to-leading-order.
Here we take the decay $B^0\to \psi(2S)K^0$ as an example, the value of
the factorizable color-suppressed amplitude $F^{V-A}_{\psi K}(a_2)$ is
about $-3.55\times 10^{-2}$, which will be partly canceled by the real
part of nonfactorizable one $Re(M^{V-A}_{\psi
K}(C_2))=1.16\times10^{-2}$. And the imagine part $Im(M^{V-A}_{\psi
K})$ is small and only about $-5.59\times10^{-3}$. When the vertex
corrections are included, the factorizable color-suppressed
amplitude $F^{V-A}_{\psi K}(a_2)$ becomes a complex number. It's
real part reduces to $-1.49\times10^{-2}$, which will be largely
canceled by $Re(M^{V-A}_{\psi
K}(C_2))=1.09\times10^{-2}$ (the difference from the leading order
value is because of using the NLO Wilson coefficient). While the
imagine part induced by the vertex corrections is large and
about $-3.42\times10^{-2}$. So the total amplitude from the tree
operators increases after including the NLO contributions. From the
numerical results, we find that the contribution (the square of the
amplitudes) from the penguin operators is very small compared with
that from the tree operators, about $1.2\%$ at the leading order and
$0.05\%$ at the next-to-leading order. There is the similar
situation in the decay $B^0\to J/\Psi K^0$ \cite{li0,Grossman}. That
is to say the penguin pollution is very small in these decays. From our results, we
can find that the branching ratios of the channels $B\to \psi(2S)K,\eta_c(2S)K$
 are larger than those of the decays $B\to \psi(2S)\pi,\eta_c(2S)\pi$
by one even two orders. This is mainly because of the CKM suppression
factor $\lambda=0.22$ within the latter. Just like the argument given in Ref.\cite{beneke2} that
the soft gluon contribution is suppressed by a factor $\Lambda_{QCD}/(m_b\alpha_s)$ rather than
$\Lambda_{QCD}/m_b$ in this type of decay,
so the perturbative and power corrections can be sizeable. Our predictions shown in
Table \ref{tab1} support this argument: the NLO contributions can provide a $(52\sim55)\%$ enhancement
for the decays involving $\eta_c(2S)$ meson. And the enhancement for the decays involving $\psi(2S)$ meson
is much more large.
Furthermore,
the charmed meson rescattering effects \cite{Colangelo} in the decays $B\to \psi(2S)K,\eta_c(2S)K$
might not be very important. While the
branching ratios for the decays $B^+\to \psi(2S)\pi^+, B^0\to
\psi(2S)\pi^0$ are still smaller than the data even with the
vertex corrections and the NLO Wilson coefficients included. Maybe other
possible higher order contributions or the contributions from the Glauber gluons \cite{glauber}
in the spectator diagrams play the important roles.

In the following we will discuss the CP-violating asymmetries of $B\to \psi(\eta_c) \pi, \psi(\eta_c) K$ decays.
For the charged decays $B^+\to \psi(\eta_c) K^+$, there is no weak phase in their decay amplitudes, so the direct CP
asymmetries of these decays are zero, and it is in agreement with the experimental value $A_{CP}(B^+\to\psi(2S)K^+)=(-2.4\pm2.3)\%$ within
$1\sigma$ error. The direct CP asymmetries of the charged decay $B^+\to \psi(2S) \pi^+$ predicted by pQCD approach are listed as:
\be
A_{CP}(B^+\to\psi(2S)\pi^+)&=&(2.16^{+0.63+0.54+0.19+0.01}_{-0.55-1.03-0.29-0.01})\%,  \;\;\;     \text{(LO)} \\
A_{CP}(B^+\to\psi(2S)\pi^+)&=&(0.51^{+0.08+0.00+0.01+0.01}_{-0.13-0.08-0.00-0.00})\%,  \;\;\;     \text{(NLO)}
\en
which are consistent with the experimental data $(2.2\pm8.5\pm1.6)\%$ determined by Belle \cite{belle} and
$(4.8\pm9.0\pm1.1)\%$ by LHCb \cite{lhcb} with no evidence of direct CP violation being seen.
\begin{figure}[t]
\begin{center}
\vspace{-0.5cm} \centerline{\hspace{0.3cm}\epsfxsize=10.5cm
\epsffile{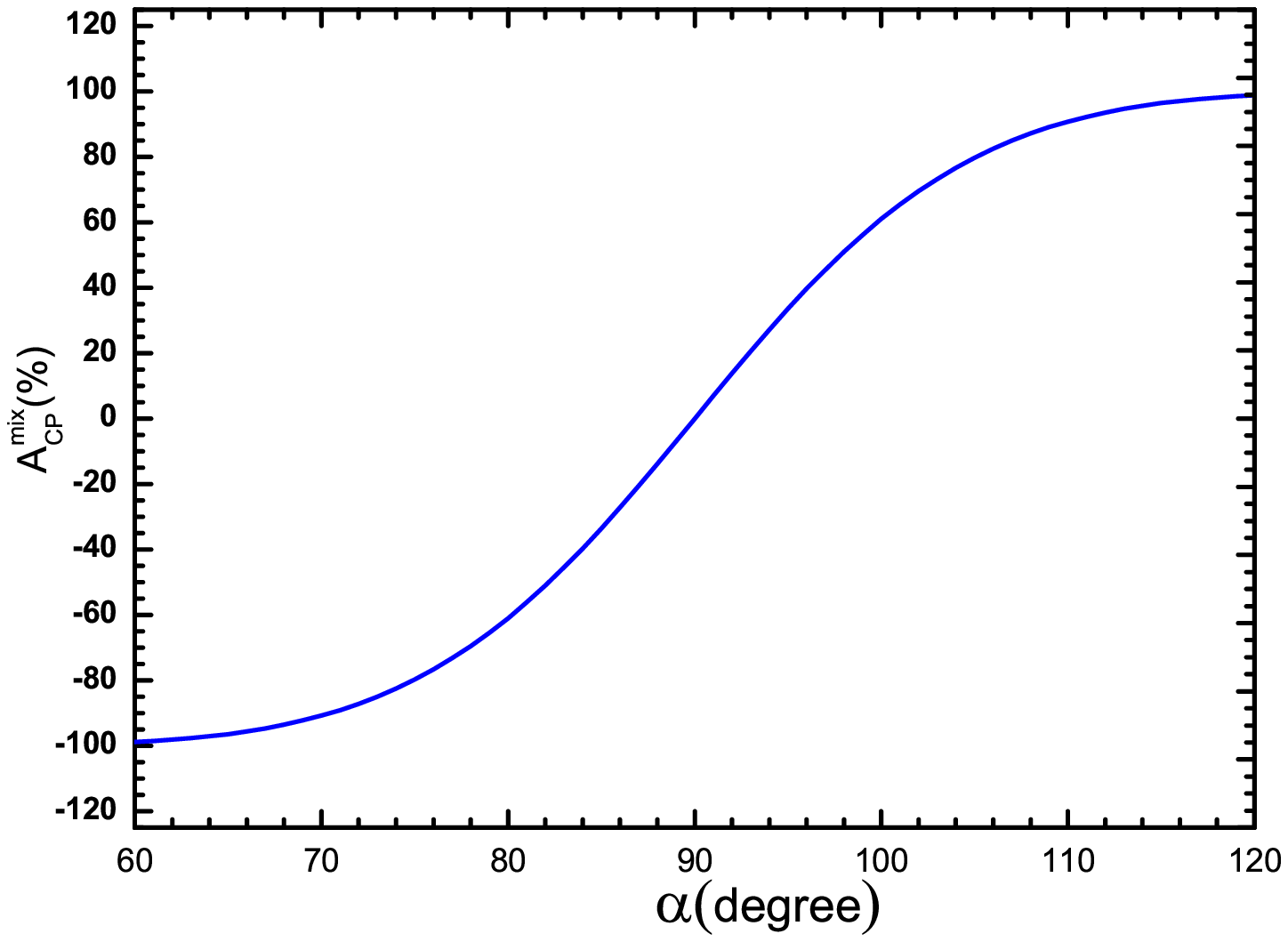} \hspace{-2.5cm} \epsfxsize=10.5cm
\epsffile{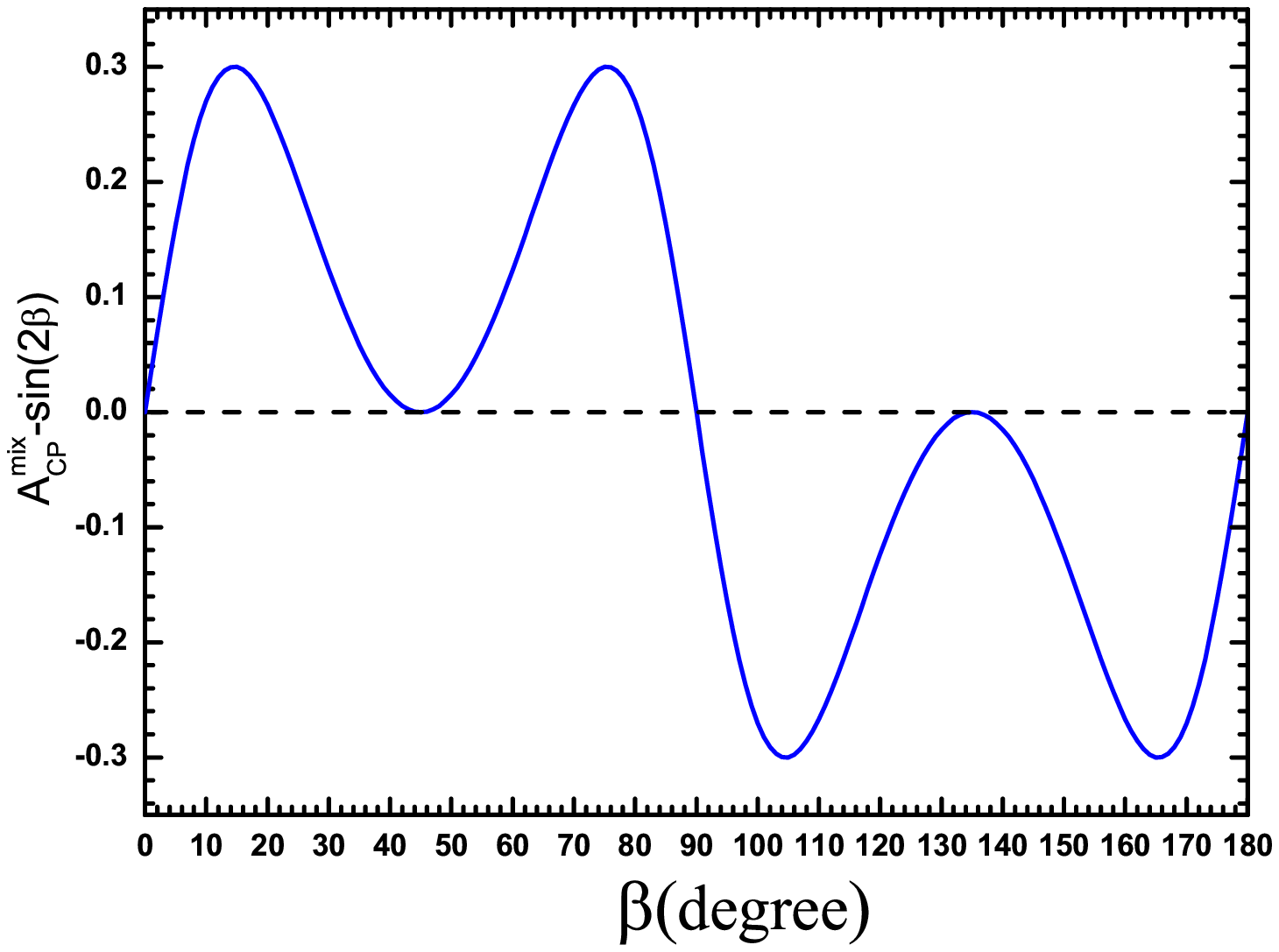}  } \vspace{-0.8cm} \caption{The mixing induced CP violating asyemtry $A^{mix}_{CP}$ and
the parameter $A^{mix}_{CP}-sin(2\beta)$
for the decays $B^0\to\psi(2s)\pi^0$ (the left) and $B^0\to\psi(2s)K^0_S$ (the rihgt), respectively.}
\label{fig3}
\end{center}
\end{figure}
As for the neutral decay channels, the effects of $B^0-\bar B^0$ mixing should be considered. The direct and mixing
induced CP violating asymmetries are defined as:
\be
A^{dir}_{CP}=\frac{|\lambda_{CP}|^2-1}{1+|\lambda_{CP}|^2},A^{mix}_{CP}=\frac{2Im(\lambda_{CP})}{1+|\lambda_{CP}|^2},
\en
where the CP-violating parameter $\lambda_{CP}$ is
\be
\lambda_{CP}=\eta_f\frac{V^*_{tb}V_{td}\langle f|H_{eff}|\bar B^0\rangle}{V_{tb}V^*_{td}\langle f|H_{eff}|B^0\rangle}=
\eta_f e^{-2i\alpha(\beta)}\frac{\langle f|H_{eff}|\bar B^0\rangle}{\langle f|H_{eff}|B^0\rangle},
\en
for $b\to d (b\to s)$ transition and $\eta_f$ is the CP-eigenvalue of the final states.
From the left panel in Fig.{\ref{fig3}}, one can find that the dependence of the mixing CP-asymmetry $A^{mix}_{CP}(B^0\to\psi(2s)\pi^0)$ on
the CKM weak phase $\alpha$. If taking the CKM weak phase $\alpha=(85.4^{+3.9}_{-3.8})^\circ$ \cite{pdg14}, we find that
the value of $A^{mix}_{CP}(B^0\to\psi(2s)\pi^0)$ is $(-31.3^{+26.4}_{-22.3})\%$. The mixing induced CP violating asymmetry for
the channel $B^0\to\psi(2s)\pi^0$ is very sensitive to the angle $\alpha$. As to another CKM weak phase $\beta$, there are much
more uncertainties: by including earlier $\sin(2\beta)$ measurements \cite{exper} and recent results from LHCb \cite{lhcb2} and Belle \cite{belle2}, the
Heavy Flavor Averaging Group (HFAVG) gives two possible solutions $2\beta=(43.8\pm1.4)^\circ$ and $2\beta=(136.2\pm1.4)^\circ$ \cite{hfavg}, which
is very consistent with our predictions shown in the right panel of Fig.3.
If the assumption that $A^{dir}_{CP}=0$ is relaxed, then $A^{mix}_{CP}=-\eta_f\sqrt{1-A^{dir}_{CP}}\sin{2\beta}$.
There are similar results between the decays
$B^0\to \psi(2S) K^0_S(\pi^0)$ and $B^0\to \eta_c(2S) K^0_S(\pi^0)$ about the mixing induced CP violating asymmetries. As for the direct
CP asymmetries of the decays $B^0\to \eta_c(2S) \pi^0$ and $B^+\to \eta_c(2S) \pi^+$, they are very small and at $10^{-5}$ order.
\section{Summary}
We study the B meson decays $B\to \psi(2S)K(\pi), \eta_c(2S)K(\pi)$
within the pQCD approach, where the radially excited charmonia states are
involved. With the wave functions of these two mesons $\psi(2S)$
and $\eta_c(2S)$ derived from the harmonic-oscillator-model, we find
that the branching ratios for the decays $B^+\to
\psi(2S)K^+,\eta_c(2S)K^+$ and $B^0\to \psi(2S)K^0$ can agree well
with the data within errors after including the NLO corrections.
While there is still some room left for other high order
corrections or the non-perturbative long distance contributions for
the decays $B^+\to \psi(2S)\pi^+$ and $B^0\to \psi(2S)\pi^0$. The
pQCD predictions for the direct CP-violating asymmetries support the
present experimental opinion, that is no evidence of direct CP
violation being observed in these decays. If a few percent value is
confirmed in the future, it would indicate new physics definitely.
\section*{Acknowledgment}
This work is partly supported by the National Natural Science
Foundation of China under Grant No. 11347030, by the Program of
Science and Technology Innovation Talents in Universities of Henan
Province 14HASTIT037.

\end{document}